\documentclass[conference]{IEEEconf}

\input epsf

\usepackage{tabularx}
\usepackage{longtable}
\usepackage{graphicx}
\usepackage{multirow}
\usepackage{hyperref}
\usepackage{pifont} 
\usepackage[dvipsnames, svgnames, x11names]{xcolor}
\usepackage{colortbl}
\usepackage{booktabs}
\usepackage{multicol}
\usepackage{stfloats}
\usepackage{listings}
\usepackage{makecell}
\usepackage{listings}
\usepackage{color}
\usepackage{xcolor}
\usepackage{paralist}
\usepackage[normalem]{ulem}
\useunder{\uline}{\ul}{}
\usepackage{tcolorbox}
\definecolor{dkgreen}{rgb}{0,0.6,0}
\definecolor{gray}{rgb}{0.5,0.5,0.5}
\definecolor{mauve}{rgb}{0.58,0,0.82}
\renewcommand\thesection{\arabic{section}} 

\renewcommand\thesubsectiondis{\thesection.\arabic{subsection}}

\renewcommand\thesubsubsectiondis{\thesubsectiondis.\arabic{subsubsection}}

\begin{document}

\title{\textbf{\Large Which is a better programming assistant? A comparative study between chatgpt and stack overflow\\}}

\author{Jinrun Liu$^{*}$, Xinyu Tang$^{*}$, Linlin Li, Panpan Chen, and Yepang Liu$^{\IEEEauthorrefmark{2}}$\\
	\normalsize Southern University of Science and Technology, Shenzhen, Guangdong, China\\
	\normalsize 12011216@mail.sustech.edu.cn, 12011439@mail.sustech.edu.cn, lill3@mail.sustech.edu.cn,\\11910236@mail.sustech.edu.cn, liuyp1@sustech.edu.cn\\
	\normalsize *co-first authors 
 \normalsize \IEEEauthorrefmark{2}corresponding author
}


\maketitle
\begin{abstract}
Programmers often seek help from Q\&A websites to resolve issues they encounter during programming. Stack Overflow has been a widely used platform for this purpose for over a decade. Recently, revolutionary AI-powered platforms like ChatGPT have quickly gained popularity among programmers for their efficient and personalized programming assistance via natural language interactions. Both platforms can offer valuable assistance to programmers, but it's unclear which is more effective at enhancing programmer productivity. In our paper, we conducted an exploratory user study to compare the performance of Stack Overflow and ChatGPT in enhancing programmer productivity. Two groups of students with similar programming abilities were instructed to use the two platforms to solve three different types of programming tasks: algorithmic challenges, library usage, and debugging. During the experiments, we measured and compared the quality of code produced and the time taken to complete tasks for the two groups. The results show that, concerning code quality, ChatGPT outperforms Stack Overflow significantly in helping complete algorithmic and library-related tasks, while Stack Overflow is better for debugging tasks. Regarding task completion speed, the ChatGPT group is obviously faster than the Stack Overflow group in the algorithmic challenge, but the two groups have a similar performance in the other two tasks. Additionally, we conducted a post-experiment survey with the participants to understand how the platforms have helped them complete the programming tasks. We analyzed the questionnaires to summarize ChatGPT and Stack Overflow's strengths and weaknesses pointed out by the participants. By comparing these, we identified the reasons behind the two platforms' divergent performances in programming assistance.

\end{abstract}
\IEEEoverridecommandlockouts
\vspace{1.5ex}

%
\IEEEpeerreviewmaketitle

\section{Introduction}


Programmers often encounter situations where they need help and guidance to complete their tasks. This is where question-and-answer (Q\&A) websites come into play. These websites provide a platform for programmers to ask questions and receive answers from other programmers who have faced similar challenges. Some of the most popular Q\&A websites include Stack Overflow~\cite{Stack_Overflow}, GitHub~\cite{GitHub}, and programming-related subreddits hosted by Reddit~\cite{Reddit}. Among these websites, Stack Overflow stands out as the most widely used platform with millions of active users worldwide~\cite{stackexchange}. It has over 21 million registered users and over 5.5 million visits per day~\cite{stackexchange}. However, the recent emergence of AI copilots, such as ChatGPT~\cite{Chat_GPT}, has gained 100 million active users in just two months~\cite{maurergeo}. ChatGPT is an AI language model developed by OpenAI that can assist programmers by providing assistance, suggestions and even generating code snippets.





There are various existing research focused on user studies of Stack Overflow
~\cite{SOAsPlugin, SOLimit, badge, polite, SODataMining}, and user studies of language models for programming assistance~\cite{perry2022users, sandoval2023lost, CopilotPython, pair-program, Productivity}. However, there are very few articles comparing ChatGPT and Stack Overflow. Delile et al.~\cite{delile2023evaluating} conducted a comparative analysis between the responses provided by Stack Overflow users and the responses generated by ChatGPT for the extracted questions. 
Programmers often face tight deadlines, requiring them to complete tasks efficiently and effectively. Both Stack Overflow, a widely used traditional developer Q\&A website, and ChatGPT, a new-generation AI-powered Q\&A platform, can offer valuable assistance to programmers. 
However, there is currently no existing work that studies which platform is more effective in helping enhance programmer productivity.
When considering productivity, we take into account two factors: the quality of the code produced and the speed of completing tasks. Code quality is essential for programmers. Writing high-quality code with fewer bugs not only saves time and effort in fixing issues but also improves the reliability and stability of the software. Besides, the speed of completing tasks is often a critical factor in time-sensitive projects and rapid development allows programmers to deliver features quickly to meet market demands. The main objective of our paper is to compare ChatGPT and Stack Overflow in order to address the following two research questions:

\begin{itemize}
    \item 
    RQ1: Which platform is better at enhancing programmers' code quality?
    
    \item 
    RQ2: Which platform is better at improving programmers' speed of completing tasks?
\end{itemize}



We conducted an exploratory user study to answer these two research questions.
To design a controlled experiment, one direct approach is to have the same group of participants use both Q\&A platforms to perform identical tasks in a consistent environment. However, as tasks or task types are one-time occurrences, independent replication is not feasible. So our idea is to let two groups of participants with similar programming abilities using ChatGPT and Stack Overflow to solve three different types of programming tasks: algorithmic challenges, library usage, and debugging. Our experiment involved 44 participants, with 23 participants in the ChatGPT group and 21 participants in the Stack Overflow group. We then compare the two platforms by respectively measuring the quality of code the two groups produced and the time they spent to complete the tasks. To ensure that the two groups had similar coding abilities, we conducted a pre-experiment questionnaire to assess the coding skills and experience of the participants and carefully considered the responses to form the groups. During the experiment, to control variables, we ensured that both groups of participants completed the tasks within the same software and hardware environment and provided them with the same instructional documentation for both platforms, adhering to identical time limits. We recorded participants' actions throughout the experiment by capturing the screens. Finally, the code quality and task completion time were assessed using the objective evaluation criteria to compare the two platforms.




To gain the experiences of participants while using these two platforms, we distributed a post-experiment questionnaire and collected their responses. This questionnaire contains a rating item for ChatGPT (or Stack Overflow), as well as an open-ended question section about its strengths and weaknesses. We extracted key points from the questionnaire responses, categorized them, and arranged them in order of frequency to provide insights and answer the following research question.

\begin{itemize}
    \item 
    RQ3: What are the underlying reasons for the different performances of ChatGPT and Stack Overflow in assisting programmers?
\end{itemize}


In terms of code quality, the experimental results demonstrate that the average code scores of the ChatGPT group are significantly higher than those of the Stack Overflow
group in algorithm tasks (38.7 vs. 5.0) and library tasks (56.2 vs. 27.4). 
However, in debugging tasks, the ChatGPT group scored lower than the Stack Overflow group (65.9 vs. 84.2). To assess task completion speed, for algorithm tasks, we recorded time at three points: first successful code execution, first test case passed and submission. For library tasks, we additionally observed the time of successful library installation. For debugging tasks, we observed the time taken to fix each individual bug. By comparing these time points, we observed that the ChatGPT group is obviously faster than the Stack Overflow group in the algorithmic challenge, but the two groups have a similar performance in the other two tasks. Through the analysis of post-experiment survey responses, we found that ChatGPT is obviously better than Stack Overflow in algorithm and library tasks due to the fact that
ChatGPT can quickly generate code and provide ideas, while
Stack Overflow lacks task-related questions and answers.
Stack Overflow has an advantage in debugging tasks due
to its expertise in solving explicit exceptions and providing
helpful links.



In summary, our work makes three major contributions:
\begin{itemize}
    \item 
    To the best of our knowledge, we conducted the first exploratory user study to compare the performance of ChatGPT and Stack Overflow in enhancing programmer productivity via both quantitative and qualitative analyses. Our findings provide valuable insights into the effectiveness of these two programming assistance tools and can guide programmers in choosing the most suitable platform for their needs.
    
    \item 
    We conducted a post-experiment survey to understand participants' perspectives on both platforms. By analyzing these responses, we identified the underlying reasons behind the two platforms’ divergent performances in programming assistance.

    \item 
    We summarized ChatGPT and Stack Overflow’s strengths and weaknesses pointed out by the participants. The results may inspire future research to improve the performance of programming assistance tools and advance the state of the art in this field.

\end{itemize}

\section{Background}
\subsection{ChatGPT}

OpenAI~\cite{openai} has published the GPT series since 2018, which is used to generate text. GPT-3~\cite{GPT3} gained attention for its impressive language generation capabilities. However, it had limitations when it came to conversational interactions. To address this, OpenAI developed ChatGPT~\cite{chatgpt} by fine-tuning GPT-3 with a conversational dataset and reinforcement learning from human feedback (RLHF)~\cite{RLHF}. This makes ChatGPT particularly effective for conversational interactions.



Benefit from the broad knowledge base, ChatGPT offers several key features for programmers: (1) ChatGPT can generate code snippets based on the given requirements; (2) It can assist in error debugging by identifying common mistakes or potential issues; (3) It can offer guidance for enhancing the algorithm by optimizing code efficiency; (4) It can help with utilizing specific APIs or libraries; (5) It can offer advice on coding bets practices and design patterns.

ChatGPT has some limitations. ChatGPT lacks network search and has limited inference capability. As a result, it may give incorrect answers in areas without extensive corpus training, making its responses untrustworthy at times~\cite{easytrick}.
Furthermore, ChatGPT's responses are sensitive to the wording of the user. If a user asks a question in a different way, ChatGPT may give a different answer or not answer at all. ChatGPT's responses tend to be too long because of its over-optimization. Additionally, even if a developer designs an algorithm to prevent ChatGPT from learning harmful content, it is still possible for ChatGPT to respond inappropriately to harmful instructions.

\subsection{Stack Overflow}

Stack Overflow~\cite{spolsky2008introducing} is a popular technical question-and-answer community that provides a platform for programmers to exchange information and solve technical problems. Created by Joel Spolsky and Jeff Atwood in 2008, it has become an essential part of the global programmer community.

The main function of Stack Overflow is to provide a platform for programmers to ask technical questions and seek solutions from other members of the community. Other users can answer, vote, and edit the answers, eventually forming a widely recognized best answer. In addition, Stack Overflow also provides a range of useful features, such as user profiles, rankings, tags, search, and more, to help users find useful information and establish personal reputations.


\section{Approach}




Our approach can be divided into two parts: a comparative experiment and a post-experiment survey. The comparative experiment aims to quantitatively compare the effectiveness of ChatGPT and Stack Overflow in enhancing programmer productivity. The post-experiment survey is conducted to identify the underlying reasons for the divergent performances of these two platforms via qualitative analysis.

\subsection{The Comparative Experiment}
Our experiment involves two groups with similar programming abilities completing three types of tasks using ChatGPT and Stack Overflow, respectively, in the same environment. We then measure and compare the code quality and completion speed of these two groups. Before conducting the experiments, we administered a pre-experiment questionnaire to assess the coding skills and experience of the participants. In the following, we will first describe the six tasks in detail. Then, we will explain the pre-experiment questionnaire design, recruitment and grouping process for participants. Finally, we will discuss the experimental procedure.

\subsubsection{Tasks Used in the Experiment}


Implementing algorithms~\cite{algorithm}, calling library functions~\cite{calling_ibrary}, and debugging existing code~\cite{debug} are most common scenarios in real-life software development. As such, we have decided to design tasks that cover these three scenarios.
The quality of an algorithm has a significant impact on
the performance and efficiency of software development. 
Selecting an appropriate algorithm and
optimizing it are crucial in software development. For this scenario, we selected three algorithm tasks with increasing levels of difficulty. 
Calling library functions is essential in software development. It simplifies the process and improves code maintainability and reusability.
For this scenario, we selected a typical library-related task. 
Debugging is essential for software development. It ensures the high quality and reliability of software.
We designed two debugging tasks, one involving fixing logical bugs and the other related to library utilization bugs.

We designed six tasks that correspond to the software development scenarios described above, which require the participants to complete them.
Table 1
presents the detailed information of these six tasks, including task descriptions, specific instructions (input, output, constraints), and possible solutions. Table 2 
shows the criteria we used to evaluate task completion.

\begin{table*}[]
\centerline {Table 1. Six tasks used in the comparative experiment (the numbers 1 to 6 in the table represent the names of the tasks as}
\centerline {follows: ``String matching~\cite{Q1}"; `The longest non-increasing subsequence~\cite{Q2}"; ``Number of primes~\cite{Q3}"; ``Extract text from}
\centerline {images using OpenCV~\cite{OpenCV} and Tesseract OCR libraries~\cite{Q4}"; ``Greatest rectangle area~\cite{Q5}"; ``The wine classifier~\cite{Q6}")}
\fontsize{7.8pt}{\baselineskip}\selectfont 
\tabcolsep=0.11cm
\renewcommand\arraystretch{1}
\begin{tabularx}{1.0\textwidth}{|c|c|cX|}
\Xhline{1.8px}
\textbf{Types}                            & \textbf{Tasks}               & \multicolumn{2}{c|}{\textbf{Contents}}                                                                                                                                                                                                                                                                                                                                                                                                                                                                                            \\ \Xhline{1.5px}
                                     &                              & \multicolumn{1}{c|}{\textit{Description:}}                       & Given a template string of length n and k strings of length up to L, find the string that has the longest common substring with the template string, and find the length of that longest common string                                                                                                                                                                                                                                                         \\ \cline{3-4} 
                                     &                              & \multicolumn{1}{c|}{\textit{Instructions:}}                      & \textbf{Input:} The first line is the template string. The second line is the value of $k$. From the third line to the $(k+2)$th line, each line has a string.  \textbf{Output:} The length of the longest common substring.  \textbf{Constraints:} (1) Every string consists of only lowercase English letters; (2) $1 \leq k \leq 10^9$; 3. $1 \leq  L  \leq 10^2$ (L is the length of each string);                                                                                       \\ \cline{3-4} 
                                     & \multirow{-3}{*}{\textbf{1}} & \multicolumn{1}{c|}{\cellcolor[HTML]{EFEFEF}\textit{Solutions:}} & \cellcolor[HTML]{EFEFEF}(1) Brute Force: $\mathcal O(L*n*k)$; (2) KMP: $\mathcal O((L+n)*k)$;                                                                                                                                                                                                                                                                                                                                                                  \\ \cline{2-4} 
                                     &                              & \multicolumn{1}{c|}{\textit{Description:}}                       & Given an array, find the longest non-increasing subsequence and the longest increasing subsequence in the sequence.                                                                                                                                                                                                                                                                                                                                            \\ \cline{3-4} 
                                     &                              & \multicolumn{1}{c|}{\textit{Instructions:}}                      & \textbf{Input:} A sequence composed with integers. \textbf{Output:} The first line is the length of the longest non-increasing subsequence and the second line is the length of longest increasing subsequence. \textbf{Constrains:} (1) For the first half of the evaluation data, the length of array is less than $10^4$; (2) For the rest evaluation data, the length of array is less than $10^5$; (3) For all data, the length of array is a integer and less than $5 \times 10^4$; \\ \cline{3-4} 
                                     & \multirow{-3}{*}{\textbf{2}} & \multicolumn{1}{c|}{\cellcolor[HTML]{EFEFEF}\textit{Solutions:}} & \cellcolor[HTML]{EFEFEF}(1) Dynamic Programming: $\mathcal O(n^2)$; (2) Dynamic Programming plus Binary Search: $\mathcal O(nlogn)$;                                                                                                                                                                                                                                                                                                                           \\ \cline{2-4} 
                                     &                              & \multicolumn{1}{c|}{\textit{Description:}}                       & Given an integer n, calculate the value of $\pi$ (n). $\pi$(n) represents the number of primes in the range 1 to n. Please use a more efficient algorithm as much as possible.                                                                                                                                                                                                                                                                                 \\ \cline{3-4} 
                                     &                              & \multicolumn{1}{c|}{\textit{Instructions:}}                      & \textbf{Input:} A integer $n$. \textbf{Output:} The number of primes between 1 to $n$. \textbf{Constrains:} $n \leq 10^{13}$.                                                                                                                                                                                                                                                                                                                                                             \\ \cline{3-4} 
\multirow{-16}{*}{\textbf{Algorithm}} & \multirow{-3}{*}{\textbf{3}} & \multicolumn{1}{c|}{\cellcolor[HTML]{EFEFEF}\textit{Solutions:}} & \cellcolor[HTML]{EFEFEF}(1) Brute Force: $\mathcal O(n^2)$; (2) Eratosthenes: $\mathcal O(nloglogn)$; (3) Eyler: $\mathcal O(n)$; (4) Meissel-Lehmer: $\mathcal O(n^{2/3}/log^2n)$;                                                                                                                                                                                                                                                                            \\ \Xhline{1.5px}
                                     &                              & \multicolumn{1}{c|}{\textit{Description:}}                       & Write a program that aims to extract text from images as accurately as possible.                                                                                                                                                                                                                                                                                                                                                                               \\ \cline{3-4} 
                                     &                              & \multicolumn{1}{c|}{\textit{Instructions:}}                      &  \textbf{Input:} Images including text.  \textbf{Output:} The recognized text.  \textbf{Constraints:} All the text used for testing is either in English or in Chinese.                                                                                                                                                                                                                                                                                                                       \\ \cline{3-4} 
\multirow{-3}{*}{\textbf{Library}}   & \multirow{-3}{*}{\textbf{4}} & \multicolumn{1}{c|}{\cellcolor[HTML]{EFEFEF}\textit{Solutions:}} & \cellcolor[HTML]{EFEFEF}Implement image recognition using OpenCV and Tesseract OCR libraries, which can extract text from images.                                                                                                                                                                                                                                                                                                                              \\ \Xhline{1.5px}
                                     &                              & \multicolumn{1}{c|}{\textit{Description:}}                       & Given an array of integers `heights` representing the histogram's bar height where the width of each bar is `1`, return the area of the largest rectangle in the histogram. We provided a code with errors, please find the errors in the code and correct them.                                                                                                                                                                                               \\ \cline{3-4} 
                                     &                              & \multicolumn{1}{c|}{\textit{Instructions:}}                      &  \textbf{Input:} An array of integers representing the heights.  \textbf{Output:} The number of areas of the largest rectangle in the histogram.  \textbf{Constraints:} (1) $1 \leq heights.length \leq 10^5$; (2) $ 0 \leq heights[i] \leq 10^4 $;                                                                                                                                                                                                                                            \\ \cline{3-4} 
                                     & \multirow{-3}{*}{\textbf{5}} & \multicolumn{1}{c|}{\cellcolor[HTML]{EFEFEF}\textit{Solutions:}} & \cellcolor[HTML]{EFEFEF}Analyse the code we have provided and fix the three logical errors in it to implement the complete monotonic stack algorithm.                                                                                                                                                                                                                                                                                                          \\ \cline{2-4} 
                                     &                              & \multicolumn{1}{c|}{\textit{Description:}}                       & Given the dataset data.csv, you are asked to modify the code to implement the J48 classifier to predict the quality of wine.                                                                                                                                                                                                                                                                                                                                   \\ \cline{3-4} 
                                     &                              & \multicolumn{1}{c|}{\textit{Instructions:}}                      &  \textbf{Input:} A CSV file, with wine's features (fixed acidity, volatile acidity, citric acid, residual sugar, chlorides, free sulfur dioxide, total sulfur dioxide, density, pH, sulfates, alcohol, quality).  \textbf{Output:} The confusion matrix, accuracy, and the size of training and testing sets.  \textbf{Constraints:} The training and testing sets should be delimited by yourself, in a ratio of 8:2.                                                                            \\ \cline{3-4} 
\multirow{-6}{*}{\textbf{Debugging}} & \multirow{-3}{*}{\textbf{6}} & \multicolumn{1}{c|}{\cellcolor[HTML]{EFEFEF}\textit{Solutions:}} & \cellcolor[HTML]{EFEFEF}Analyse the code we have provided and fix three of the running errors and two of the logic errors to implement the J48 classifier. In addition, the final accuracy of the classifier should be more than 85\% to prove that the code was correctly corrected.                                                                                                                                                                          \\ \Xhline{1.8px}
\end{tabularx}
\label{six_tasks}
\end{table*}



\begin{table*}
\centerline{Table 2. Evaluation criteria for the six tasks used in the comparative experiment}
\fontsize{8.2pt}{\baselineskip}\selectfont 
\tabcolsep=0.3cm
\renewcommand\arraystretch{1}
\begin{tabular}{|c|p{15.5cm}|}
\Xhline{1.5px}
\multicolumn{1}{|c|}{\textbf{Types}}       & \multicolumn{1}{c|}{\textbf{Evaluation}}                                                                                                                                                                                                                                                                                                                                                                                                                                                                                                                                                                                                                                                                                                                                                                                                                                           \\ \Xhline{1.5px}
\textbf{Algorithm} & For each task, the full score is one hundred points. For task 1, there are 10 test cases in total, and passing one can obtain ten points. For task 2 and 3, there are 20 test cases for each task, with each test case worth 5 points. For all three tasks, the time limit for each testing point is 1s, and the space limit is 128 MB.                                                                                                                                                                                                                                                                                                                                                                                                                                                                                                 \\ \hline
\textbf{Library}   & The scoring of this task is divided into two parts, with a total score of one hundred points. The first part accounts for 10 points, and participants can earn these points by successfully installing the corresponding library. The second part accounts for 90 points and consists of 8 test cases, each corresponding to an image. A total of 494 English or Chinese characters need to be abstracted. Assuming that the participant's provided code correctly recognizes $x$ characters, the score for this part of the task would be $\frac{x}{494}\times90$ points. The final score is the sum of the scores for both parts.                                                                                                                                                                                                        \\ \hline
\textbf{Debugging} & In task 5, we provide erroneous code that contains five bugs that need to be fixed. These bugs consist of one runtime error and four logical errors within the algorithm. The following is a list of all the bugs, where the first is the runtime error and the last four are logical errors: 1) Empty stack judgment is missing; 2) Avoid division by 0; 3) The stack needs to be cleared; 4) Empty stack judgment is missing; 5) Avoid division by 0; In task 6, in the code provided, there are a total of six bugs, including four runtime errors and two logical errors. The following is a list of all the bugs, where the first four are runtime errors and the last two are logical errors. After fixing the first four bugs, the code can run successfully but produces incorrect results that do not meet the requirements: 1) The file path is incorrect; 2) The class attribute is wrong; 3) The filter should be NumericToNominal instead of StringToNomial; 4) Data has not been divided into 3 classes as the quality; 5) The size of the training set and test set is not 8:2; 6) The accuracy is less than 85\%. \\  \Xhline{1.5px}
\end{tabular}
\label{evaluation}
\end{table*}

\subsubsection{The Pre-experiment Questionnaire}

\begin{table*}[ht]

\centerline{Table 3. Pre-experiment questionnaire}
\fontsize{8.9pt}{\baselineskip}\selectfont 

  \begin{tabular}{l p{13cm} l}
			\toprule[2pt]
\textbf{\small{\#}} & \textbf{\small{Questions}} & \textbf{\small{Notes}}
            \vspace{1mm}
			\\
			\midrule[1pt]
			\rowcolor{WhiteSmoke} 1 & What is your name? & Basic information
			\vspace{1.5mm}
			\\
		2 & What is your grade?  & Basic information
			\vspace{1.5mm}
			\\
			\rowcolor{WhiteSmoke} 3 &	Have you completed the following courses: Java (CS102A/CS109), DSAA (CS203), Algorithm Design (CS208), and AI (CS303)? If yes, please provide your grades for each course.
   & Basic information
			\vspace{1.5mm}
			\\
			4 & How many years of programming experience do you have? & Programming experience
			\vspace{1.5mm}
			\\
		\rowcolor{WhiteSmoke} 5 & 
  Please provide a list of programming languages you are familiar with, ranked in order of familiarity.
  & Programming languages
			\vspace{1.5mm}
			\\
			6 & How many open-source projects have you contributed to?  & Project impact
			\vspace{1.5mm}
			\\
		\rowcolor{WhiteSmoke} 7 & How many followers do you have on Github?  &  Project impact
			\vspace{1.5mm}
			\\
			8 & How many stars do you have in total on Github?  & Project impact
			\vspace{1.5mm}
			\\
   \rowcolor{WhiteSmoke}	9 & Have you ever used maven to import a library in Java? Please rate your familiarity on a scale of 1-5.  & Specific programming skills
			\vspace{1.5mm}
			\\
			\bottomrule[2pt]
		\end{tabular}
\label{pre}
\end{table*}


There are two purposes for conducting a questionnaire survey to assess the programming ability of participants: 1) to assign tasks to these participants, ensuring that they have the required knowledge and ability to complete the tasks assigned to them; 2) to ensure that the ChatGPT and Stack Overflow groups assigned the same task have similar programming abilities. 

The questionnaire was designed to collect basic information and assess the programming proficiency of the participants. As shown in Table 3, it consisted of nine questions covering five aspects: basic information, programming experience, programming languages, project impact, and a specific programming skill. Q1-Q3 were employed to collect basic information, including name, grade, as well as completion status and grades of four courses: Java Programming (JAVA), Data Structure and Algorithm Analysis (DSAA), Algorithm Design, and Artificial Intelligence (AI), which were related to our tasks. Q4 gathered participants' number of years of programming experience, while Q5 inquired about the programming languages they were familiar with and their proficiency level. Q6-Q8 collected information on the number of projects, followers, and stars on participants' GitHub profiles, helping us understand their project impact. Q9 aimed to assess participants' familiarity with Maven~\cite{Maven}, which will help us determine the participants for task 4, which is related to Maven. Notice that 
our questionnaire was designed according to the guidelines outlined in \textit{Guidelines for Conducting Surveys in Software Engineering v. 1.1}~\cite{linaker2015guidelines} to ensure its reliability.


\subsubsection{Recruitment and Grouping of Participants}


To recruit participants interested in completing programming tasks using ChatGPT and Stack Overflow, we offered prizes and designed an advertisement for the experiment. The ads were placed on social media platforms to attract a sufficient number of individuals to sign up for the experiment. Finally, we received 44 registrations. In the following, we will first describe how we assigned tasks to the participants and then introduce how they were assigned to either the ChatGPT or Stack Overflow subgroup. 

Since Task 4 and Task 6 not only require participants to have programming experience but also have specific programming skills, Task 4 requires participants to have proficiency in using Maven, while Task 6 requires participants to have the ability to train neural network models, we will first find suitable participants for these two tasks. Assignment of participants to Task 4 is determined by their response to Q9. If they have not used Maven, they will not be assigned to Task 4. A total of 21 participants are eligible to be assigned to this task. Assignment of participants to Task 6 is determined by their response to Q2. Participants who are less than a junior, indicating that they have not studied artificial intelligence courses, cannot be assigned to Task 6. There are 13 participants who are eligible to be assigned to this task. Following a discussion between three researchers from our team, we selected 6 eligible participants with comparable programming abilities to complete Task 6.
Programming abilities were obtained by analyzing their responses to Q3-Q8 in the questionnaire. From the remaining 16 eligible participants, we then selected another 6 with comparable abilities to complete Task 4.

All the remaining participants were assigned to the algorithm-related Task 1, 2, 3 and 5. After a discussion among three researchers from our team, we ranked the four tasks by difficulty level, from easiest to hardest, as follows: Task 1, Task 2, Task 5, and Task 3. To ensure that the remaining participants were assigned appropriate tasks, we further discussed their responses to Q3-Q8, and ranked their programming abilities. Finally, we allocated tasks based on the principle of assigning more challenging tasks to participants with higher abilities. As a result, we assigned 10 participants to Task 1, 9 to Task 2, 7 to Task 3, and 6 to Task 5. Next, we will introduce how participants for each task were divided into the ChatGPT group and the Stack Overflow group. Since all participants had used both tools and were equally familiar with each of them, we divided each group equally into ChatGPT and Stack Overflow subgroups through a randomized process. Table 4 displays the final result of grouping the 44 participants into 6 tasks.


\begin{table}[]
\centerline{Table 4. Grouping of 44 participants into 6 tasks}
\tabcolsep=0.41cm
\fontsize{8pt}{\baselineskip}\selectfont 
\renewcommand\arraystretch{1}
\begin{tabular}{|c|c|c|lc|}
\Xhline{1.5px}
\textbf{Types}                      & \textbf{Tasks}              & \textbf{Total}      & \multicolumn{2}{c|}{\textbf{Group}}     \\ \Xhline{1.5px}
\multirow{6}{*}{\textbf{Algorithm}} & \multirow{2}{*}{\textbf{1}} & \multirow{2}{*}{10} & \multicolumn{1}{l|}{ChatGPT}        & 5 \\ \cline{4-5} 
                                    &                             &                     & \multicolumn{1}{l|}{Stack Overflow} & 5 \\ \cline{2-5} 
                                    & \multirow{2}{*}{\textbf{2}} & \multirow{2}{*}{9}  & \multicolumn{1}{l|}{ChatGPT}        & 5 \\ \cline{4-5} 
                                    &                             &                     & \multicolumn{1}{l|}{Stack Overflow} & 4 \\ \cline{2-5} 
                                    & \multirow{2}{*}{\textbf{3}} & \multirow{2}{*}{6}  & \multicolumn{1}{l|}{ChatGPT}        & 3 \\ \cline{4-5} 
                                    &                             &                     & \multicolumn{1}{l|}{Stack Overflow} & 3 \\ \Xhline{1.5px}
\multirow{2}{*}{\textbf{Library}}   & \multirow{2}{*}{\textbf{4}} & \multirow{2}{*}{6}  & \multicolumn{1}{l|}{ChatGPT}        & 3 \\ \cline{4-5} 
                                    &                             &                     & \multicolumn{1}{l|}{Stack Overflow} & 3 \\ \Xhline{1.5px}
\multirow{4}{*}{\textbf{Debugging}} & \multirow{2}{*}{\textbf{5}} & \multirow{2}{*}{7}  & \multicolumn{1}{l|}{ChatGPT}        & 4 \\ \cline{4-5} 
                                    &                             &                     & \multicolumn{1}{l|}{Stack Overflow} & 3 \\ \cline{2-5} 
                                    & \multirow{2}{*}{\textbf{6}} & \multirow{2}{*}{6}  & \multicolumn{1}{l|}{ChatGPT}        & 3 \\ \cline{4-5} 
                                    &                             &                     & \multicolumn{1}{l|}{Stack Overflow} & 3 \\ \Xhline{1.5px}
\end{tabular}

\label{group-table}
\end{table}

\subsubsection{Experimental Procedure}


During the experiment, each participant was provided with the same software and hardware environment for each task. Participants completed the tasks on the same computer using IntelliJ IDEA~\cite{IDEA} as the development environment. Additionally, participants were given identical task descriptions, tool usage manuals, and initial code. We ensured that all participants completed the tasks within a limited time of 1 hour without seeking assistance from any platform other than ChatGPT or Stack Overflow. Throughout the experiment, we recorded the participants' computer screens to monitor compliance with our experimental guidelines. The recorded videos also provided us with the timing of when participants completed the tasks.


\subsection{The Post-experiment Questionnaire}

\begin{table*}[h]
\centerline{Table 5. Post-experiment questionnaire}
\fontsize{9pt}{\baselineskip}\selectfont 
\tabcolsep=0.37cm


		\begin{tabular}{l p{14cm} l}
			\toprule[2pt]
	\textbf{\small{\#}} &		\textbf{\small{Questions}} & \textbf{\small{Notes}}
            \vspace{0.5mm}

        \\
                    	\midrule[1pt]

			
	
	\multirow{2}{*}{1} &  \multirow{2}{0.7\textwidth}{How helpful do you think ChatGPT (or Stack Overflow) has been for you?\\ (1. Not helpful, 2. Somewhat helpful, 3. Moderately helpful, 4. Quite helpful, 5. Very helpful)} &
       \multirow{2}{*}{Rating}
    
			\vspace{5.5mm}
   
			\\
    2 &Please specify in which ways ChatGPT (or Stack Overflow) has been helpful for you? (short answer question) & Strengths
			\vspace{1.5mm}
   \\

    3 & Please specify in what aspects ChatGPT (or Stack Overflow) is not satisfactory? (short answer question) & Weaknesses
			\vspace{1.5mm}
   \\
 
			\bottomrule[2pt]
		\end{tabular}
\label{post}
\end{table*}


The post-experiment questionnaire aims to investigate the underlying reasons for the differential effects of ChatGPT and Stack Overflow in assisting programming. The questionnaire consists of three questions, one multiple-choice question and two open-ended questions, as shown in Table 5. 
Q1 asks participants to rate the level of assistance provided by ChatGPT or Stack Overflow in completing their tasks, on a scale from 1 to 5.
Q2 and Q3 ask participants to describe the specific ways in which ChatGPT or Stack Overflow has been helpful to them and to identify any aspects in which they were not satisfactory. By analyzing the answers to Q1, we can obtain a direct comparison between ChatGPT and Stack Overflow in terms of assistance provided. Through the analysis of participants' responses to Q2 and Q3, we can gather their perceptions of the strengths and weaknesses of the two platforms.


To minimize the likelihood of participants forgetting experiment details and not being able to provide a comprehensive evaluation of ChatGPT or Stack Overflow, we administered the questionnaire survey immediately after they completed the programming tasks. The questionnaire was not subject to any time restrictions. After completing the questionnaire, the experiment was considered concluded, and the participants were free to leave the experiment location.

\section{Results}

\subsection{RQ1: Comparison of Code Quality}

\begin{table*}[]
\centerline{Table 6. Comparison of code quality between ChatGPT group and Stack Overflow group}
\centering
\fontsize{8.5pt}{\baselineskip}\selectfont 
\tabcolsep=0.37cm
\renewcommand\arraystretch{1.2}
\begin{tabular}{|c|c|cccccccccccc|}
\Xhline{1.8px}
                                     &                                  & \multicolumn{12}{c|}{\textbf{Scores (out of 100)}}                                                                                                                                                                                                                                                                                                                                                                                                                                                                                   \\ \cline{3-14} 
\multirow{-2}{*}{\textbf{Types}}     & \multirow{-2}{*}{\textbf{Tasks}} & \multicolumn{5}{c|}{\textbf{ChatGPT group}}                                                                                                                    & \multicolumn{1}{c|}{\cellcolor[HTML]{DADADA}\textit{\textbf{Avg}}} & \multicolumn{5}{c|}{\textbf{Stack Overflow group}}                                                                                                                                                                                             & \cellcolor[HTML]{DADADA}\textit{\textbf{Avg}} \\ \Xhline{1.5px}
                                     & \textbf{1}                       & \multicolumn{1}{c|}{0}                           & \multicolumn{1}{c|}{80}   & \multicolumn{1}{c|}{80}   & \multicolumn{1}{c|}{100} & \multicolumn{1}{c|}{100} & \multicolumn{1}{c|}{\cellcolor[HTML]{DADADA}72.0}                  & \multicolumn{1}{c|}{0}                              & \multicolumn{1}{c|}{0}                              & \multicolumn{1}{c|}{0}                              & \multicolumn{1}{c|}{0}                              & \multicolumn{1}{c|}{0} & \cellcolor[HTML]{DADADA}0.0                   \\ \cline{2-14} 
                                     & \textbf{2}                       & \multicolumn{1}{c|}{2}                           & \multicolumn{1}{c|}{2}    & \multicolumn{1}{c|}{5}    & \multicolumn{1}{c|}{20}  & \multicolumn{1}{c|}{50}  & \multicolumn{1}{c|}{\cellcolor[HTML]{DADADA}15.8}                  & \multicolumn{1}{c|}{{\color[HTML]{000000} {\ul 0}}} & \multicolumn{1}{c|}{{\color[HTML]{000000} {\ul 0}}} & \multicolumn{1}{c|}{{\color[HTML]{000000} {\ul 0}}} & \multicolumn{1}{c|}{{\color[HTML]{000000} {\ul 0}}} & \multicolumn{1}{c|}{}  & \cellcolor[HTML]{DADADA}0.0                   \\ \cline{2-12} \cline{14-14} 
\multirow{-3}{*}{\textbf{Algorithm}} & \textbf{3}                       & \multicolumn{1}{c|}{5}                           & \multicolumn{1}{c|}{40}   & \multicolumn{1}{c|}{40}   & \multicolumn{2}{c|}{}                               & \multicolumn{1}{c|}{\cellcolor[HTML]{DADADA}28.3}                  & \multicolumn{1}{c|}{5}                              & \multicolumn{1}{c|}{5}                              & \multicolumn{1}{c|}{35}                             & \multicolumn{2}{c|}{}                                                        & \cellcolor[HTML]{DADADA}15.0                  \\ \Xhline{1.5px}
\textbf{Library}                     & \textbf{4}                       & \multicolumn{1}{c|}{{\color[HTML]{333333} 10.0}} & \multicolumn{1}{c|}{78.1} & \multicolumn{1}{c|}{80.5} & \multicolumn{2}{c|}{}                               & \multicolumn{1}{c|}{\cellcolor[HTML]{DADADA}56.2}                  & \multicolumn{1}{c|}{{\color[HTML]{333333} {\ul 0}}} & \multicolumn{1}{c|}{{\color[HTML]{333333} {\ul 0}}} & \multicolumn{1}{c|}{82.3}                           & \multicolumn{2}{c|}{}                                                        & \cellcolor[HTML]{DADADA}27.4                  \\ \Xhline{1.5px}
                                     & \textbf{5}                       & \multicolumn{1}{c|}{0}                           & \multicolumn{1}{c|}{90}   & \multicolumn{1}{c|}{90}   & \multicolumn{1}{c|}{100} & \multicolumn{1}{c|}{}    & \multicolumn{1}{c|}{\cellcolor[HTML]{DADADA}70.0}                  & \multicolumn{1}{c|}{90}                             & \multicolumn{1}{c|}{90}                             & \multicolumn{1}{c|}{90}                             & \multicolumn{2}{c|}{}                                                        & \cellcolor[HTML]{DADADA}90.0                  \\ \cline{2-6} \cline{8-11} \cline{14-14} 
\multirow{-2}{*}{\textbf{Debugging}} & \textbf{6}                       & \multicolumn{1}{c|}{45}                          & \multicolumn{1}{c|}{60}   & \multicolumn{1}{c|}{80}   & \multicolumn{2}{c|}{}                               & \multicolumn{1}{c|}{\cellcolor[HTML]{DADADA}61.7}                  & \multicolumn{1}{c|}{55}                             & \multicolumn{1}{c|}{80}                             & \multicolumn{1}{c|}{100}                            & \multicolumn{2}{c|}{\multirow{-2}{*}{}}                                      & \cellcolor[HTML]{DADADA}78.3                  \\ \Xhline{1.8px}
\end{tabular}
\end{table*}


For the six tasks, we evaluated the code produced by 44 participants in both the ChatGPT and Stack Overflow groups using the criteria outlined in Table 2. The comparison of code quality scores between the two groups is presented in Table 6. The two ``\textit{Avg}" columns in the table display the average scores for the ChatGPT and Stack Overflow groups, respectively. Notice that the maximum score for each of six tasks was set to 100 points. In the table, a score of 0 with an underline indicates that the code submitted by the participant could not be successfully executed, whereas a score of 0 without an underline indicates that the submitted code was able to run but failed in all test cases.


Based on the analysis of the data, 
it can be concluded that for the algorithm tasks, the scores of the ChatGPT group are significantly higher than those of the Stack Overflow group (72.0 vs. 0.0; 15.8 vs. 0.0; 28.3 vs. 15.0), as well as the average scores (38.7 vs. 5.0).
In the library usage task, the ChatGPT group's score is notably higher than that of the Stack Overflow group (56.2 vs. 27.4).
However, for debugging tasks, ChatGPT's performance is inferior to that of the Stack Overflow group on Task 5 (70.0 vs. 90.0) and Task 6 (61.7 vs. 78.3). The average scores of the two groups are 65.9 and 84.2.
From this table, we have also made the following observation: all participants in the ChatGPT group successfully produced runnable code. In contrast, 28.6\% (6/21) of participants in the Stack Overflow group produced code that failed to run successfully.

\begin{tcolorbox}[boxrule=1pt,boxsep=1pt,left=2pt,right=2pt,top=2pt,bottom=2pt]
\textbf{Answer to RQ1:} 
The ChatGPT group significantly outperformed the Stack Overflow group in algorithm tasks (38.7 vs. 5.0) and library tasks (56.2 vs. 27.4). However, in debugging tasks, the ChatGPT group scored lower than the Stack Overflow group (65.9 vs. 84.2).

\end{tcolorbox}

\begin{tcolorbox}[boxrule=1pt,boxsep=1pt,left=2pt,right=2pt,top=2pt,bottom=2pt]

\textbf{Observation 1:} The ChatGPT group had a much lower rate of creating non-runnable code than the Stack Overflow group (0\% vs. 28.6\%).
 
\end{tcolorbox}


\subsection{RQ2: Comparison of Task Completion Speed}

During the experiment, we conducted screen recording of participants while they completed the tasks. After the experiment, we manually reviewed the recorded videos to determine the timestamps at which participants completed the tasks. 
For algorithm tasks, we recorded three timestamps: the time of the first successful code execution (first program execution), the time of the first successful test case passed (first test case passed), and the point of time when the code is no longer being changed (submission).
We consider the first execution of the program as an indication that the participant has a subjectively mature code that is ready for testing its correctness. Once the code passes the first test case, we consider it to have objectively matured and become relatively complete. The time of submission provides insight into the total time invested by participants in the project. We also calculated the time interval between the first program execution and the first time test case passed. This interval holds significance as it depicts the time participants took to modify the code after having a complete version in order to pass the test case. For library tasks, in addition to recording the three mentioned time points and time interval, we observed the timestamp of successful library installation. For debugging tasks, we observed the time taken to fix each individual bug.

The Comparison of timestamps for three algorithm tasks was shown in Table 7. The two ``\textit{Avg}" columns in the table display the average values for each group of participants in terms of timestamps or time intervals. It's important to note that our default start time for each task was 00:00:00, and the end time was set at 01:00:00. Task 2 has two ``first test case passed" time points, as it includes two subtasks. Furthermore, it should be noted that if a participant's code failed to run successfully during the experiment, the ``first program execution" time point was set to 01:00:00 and indicated with an underline in the table. Similarly, if a participant's code did not pass the test cases, the ``first test case passed" time point was set to 01:00:00 and also marked with an underline. During our experiment, one participant from the Stack Overflow group abandoned Task 2 midway, which resulted in the inability to accurately collect their corresponding time point data. In Table 7, we denote this with a ``-" symbol. Similarly, for cases where the "first test case passed" time point data is indicated as underlined 01:00:00, we were unable to calculate the time interval data, and in the table, we use ``-" to represent this. Table 8 shows the comparison of timestamps for the library-related task. The underlined numbers and ``-" in the table convey the same meanings as in Table 7. Table 9 shows the comparison of times spent for the two debugging tasks. Notice that one participant from the ChatGPT group abandoned Task 5, in the table, we used ``-" to indicate the relevant data for this participant.




For algorithm tasks, our analysis of Table 7 reveals that the ChatGPT group consistently precedes the Stack Overflow group at all three timestamps. Notably, the ChatGPT group achieved a clear lead in completing the first program execution ahead of the Stack Overflow group, and completed the first test case passed and submission slightly before the Stack Overflow group.  It is worth noting that we observed a significant difference in the interval between ``first program execution" and ``first test case passed" for the algorithm tasks, with the ChatGPT group having a much larger interval compared to the Stack Overflow group. This indicates that participants using ChatGPT require more time for code refinement and modifications following the successful execution of their code. For the library-related task and debugging tasks, we observed that there were no significant differences between the ChatGPT and Stack Overflow groups in terms of time spent or at various timestamps.



\begin{tcolorbox}[boxrule=1pt,boxsep=1pt,left=2pt,right=2pt,top=2pt,bottom=2pt]
\textbf{Answer to RQ2:} 
For algorithm tasks, the ChatGPT group significantly precedes the Stack Overflow group at all three timestamps. For the library-related task and debugging tasks, there were no significant differences between the ChatGPT and Stack Overflow groups at various timestamps or in terms of time spent.
\end{tcolorbox}

\begin{tcolorbox}[boxrule=1pt,boxsep=1pt,left=2pt,right=2pt,top=2pt,bottom=2pt]

\textbf{Observation 2:} The time interval between ``first program execution'' and ``first test case passed'' is notably longer for the ChatGPT group compared to the Stack Overflow group. This indicates that participants using ChatGPT require more time for code refinement and modifications following the successful execution of their code.
 
\end{tcolorbox}

\begin{table*}[!h]
\centerline{Table 7. Comparison of timestamps for three algorithm tasks}
\fontsize{7.8pt}{\baselineskip}\selectfont 
\tabcolsep=0.09cm
\renewcommand\arraystretch{1.3}

\begin{tabular}{|c|c|ccccclcccccc|}
\hline
                                 & \multicolumn{1}{c|}{}                                            & \multicolumn{12}{c|}{\textbf{Timestamps (hh:mm:ss) / Time interval (hh:mm:ss)}}                                                                                                                                                                                                                                                                                                                                                                                                                                                                                                                                                                                                                                                                                                                                                                                                                                                                                           \\ \cline{3-14} 
\multirow{-2}{*}{\textbf{Tasks}} & \multicolumn{1}{c|}{\multirow{-2}{*}{\textbf{Key process points}}}                          & \multicolumn{5}{c|}{\textbf{ChatGPT group}}                                                                                                                                                                                                                                                                                                                                                              & \multicolumn{1}{c|}{\cellcolor[HTML]{DADADA}\textit{\textbf{Avg}}}           & \multicolumn{5}{c|}{\textbf{Stack Overflow group}}                                                                                                                                                                                                                                                                                                                                         & \cellcolor[HTML]{DADADA}\textit{\textbf{Avg}}           \\ \hline
                                 & \textbf{First program execution}                                 & \multicolumn{1}{c|}{00:05:47}                                                & \multicolumn{1}{c|}{00:20:43}                                                & \multicolumn{1}{c|}{00:50:00}                                                & \multicolumn{1}{c|}{00:13:43}                                                & \multicolumn{1}{c|}{00:20:08}                                                & \multicolumn{1}{l|}{\cellcolor[HTML]{DADADA}00:22:04}                        & \multicolumn{1}{c|}{00:31:01}                                                & \multicolumn{1}{c|}{{\ul 01:00:00}}                                   & \multicolumn{1}{c|}{00:43:49}                                                & \multicolumn{1}{c|}{00:58:57}                                                & \multicolumn{1}{c|}{{\ul 01:00:00}}                                   & \multicolumn{1}{l|}{\cellcolor[HTML]{DADADA}00:50:45}   \\ \cline{2-14} 
                                 & \cellcolor[HTML]{EFEFEF}{\color[HTML]{333333} \textbf{Interval}} & \multicolumn{1}{c|}{\cellcolor[HTML]{EFEFEF}{\color[HTML]{333333} 00:54:13}} & \multicolumn{1}{c|}{\cellcolor[HTML]{EFEFEF}{\color[HTML]{333333} 00:22:56}} & \multicolumn{1}{c|}{\cellcolor[HTML]{EFEFEF}{\color[HTML]{333333} 00:10:00}} & \multicolumn{1}{c|}{\cellcolor[HTML]{EFEFEF}{\color[HTML]{333333} 00:14:25}} & \multicolumn{1}{c|}{\cellcolor[HTML]{EFEFEF}{\color[HTML]{333333} 00:03:01}} & \multicolumn{1}{c|}{\cellcolor[HTML]{DADADA}{\color[HTML]{333333} 00:24:10}} & \multicolumn{1}{c|}{\cellcolor[HTML]{EFEFEF}{\color[HTML]{333333} 00:28:59}} & \multicolumn{1}{c|}{\cellcolor[HTML]{EFEFEF}{\color[HTML]{333333} -}} & \multicolumn{1}{c|}{\cellcolor[HTML]{EFEFEF}{\color[HTML]{333333} 00:00:28}} & \multicolumn{1}{c|}{\cellcolor[HTML]{EFEFEF}{\color[HTML]{333333} 00:00:00}} & \multicolumn{1}{c|}{\cellcolor[HTML]{EFEFEF}{\color[HTML]{333333} -}} & \cellcolor[HTML]{DADADA}{\color[HTML]{333333} 00:11:47} \\ \cline{2-14} 
                                 & \textbf{First test case passed}                                    & \multicolumn{1}{c|}{{\color[HTML]{333333} {\ul 01:00:00}}}                   & \multicolumn{1}{c|}{00:43:39}                                                & \multicolumn{1}{c|}{{\ul 01:00:00}}                                          & \multicolumn{1}{c|}{00:28:08}                                                & \multicolumn{1}{c|}{00:23:09}                                                & \multicolumn{1}{l|}{\cellcolor[HTML]{DADADA}00:46:14}                        & \multicolumn{1}{c|}{{\ul 01:00:00}}                                          & \multicolumn{1}{c|}{{\ul 01:00:00}}                                   & \multicolumn{1}{c|}{00:44:17}                                                & \multicolumn{1}{c|}{00:58:57}                                                & \multicolumn{1}{c|}{{\ul 01:00:00}}                                   & \cellcolor[HTML]{DADADA}00:56:39                        \\ \cline{2-14} 
\multirow{-4}{*}{\textbf{1}}     & \textbf{Submission}                                              & \multicolumn{1}{l|}{01:00:00}                                                & \multicolumn{1}{l|}{00:47:46}                                                & \multicolumn{1}{l|}{01:00:00}                                                & \multicolumn{1}{l|}{00:41:24}                                                & \multicolumn{1}{l|}{00:29:45}                                                & \multicolumn{1}{l|}{\cellcolor[HTML]{DADADA}00:47:47}                        & \multicolumn{1}{l|}{01:00:00}                                                & \multicolumn{1}{l|}{01:00:00}                                         & \multicolumn{1}{l|}{00:45:45}                                                & \multicolumn{1}{l|}{00:59:29}                                                & \multicolumn{1}{l|}{01:00:00}                                         & \multicolumn{1}{l|}{\cellcolor[HTML]{DADADA}00:57:03}   \\ \hline
                                 & \textbf{First program execution}                                 & \multicolumn{1}{c|}{00:40:36}                                                & \multicolumn{1}{c|}{00:13:42}                                                & \multicolumn{1}{c|}{00:27:57}                                                & \multicolumn{1}{c|}{00:15:24}                                                & \multicolumn{1}{c|}{00:30:08}                                                & \multicolumn{1}{l|}{\cellcolor[HTML]{DADADA}00:24:55}                        & \multicolumn{1}{c|}{{\color[HTML]{37638D} -}}                                & \multicolumn{1}{c|}{{\ul 01:00:00}}                                   & \multicolumn{1}{c|}{00:47:36}                                                & \multicolumn{1}{c|}{00:58:54}                                                & \multicolumn{1}{c|}{}                                                 & \cellcolor[HTML]{DADADA}00:54:34                        \\ \cline{2-12} \cline{14-14} 
                                 & \cellcolor[HTML]{EFEFEF}\textbf{Interval}                        & \multicolumn{1}{c|}{\cellcolor[HTML]{EFEFEF}00:13:43}                        & \multicolumn{1}{c|}{\cellcolor[HTML]{EFEFEF}00:05:51}                        & \multicolumn{1}{c|}{\cellcolor[HTML]{EFEFEF}00:32:03}                        & \multicolumn{1}{c|}{\cellcolor[HTML]{EFEFEF}00:14:25}                        & \multicolumn{1}{c|}{\cellcolor[HTML]{EFEFEF}00:00:00}                        & \multicolumn{1}{c|}{\cellcolor[HTML]{DADADA}00:13:12}                        & \multicolumn{1}{c|}{\cellcolor[HTML]{EFEFEF}-}                               & \multicolumn{1}{c|}{\cellcolor[HTML]{EFEFEF}-}                        & \multicolumn{1}{c|}{\cellcolor[HTML]{EFEFEF}00:12:24}                        & \multicolumn{1}{c|}{\cellcolor[HTML]{EFEFEF}00:01:06}                        & \multicolumn{1}{c|}{}                                                 & \cellcolor[HTML]{DADADA}00:06:45                        \\ \cline{2-12} \cline{14-14} 
                                 &                                                                  & \multicolumn{1}{c|}{00:54:07}                                                & \multicolumn{1}{c|}{00:19:33}                                                & \multicolumn{1}{c|}{{\color[HTML]{333333} 01:00:00}}                         & \multicolumn{1}{c|}{00:29:49}                                                & \multicolumn{1}{c|}{00:30:08}                                                & \multicolumn{1}{l|}{\cellcolor[HTML]{DADADA}00:38:43}                        & \multicolumn{1}{c|}{{\color[HTML]{37638D} -}}                                & \multicolumn{1}{c|}{{\ul 01:00:00}}                                   & \multicolumn{1}{c|}{{\ul 01:00:00}}                                          & \multicolumn{1}{c|}{{\ul 01:00:00}}                                          & \multicolumn{1}{c|}{}                                                 & \cellcolor[HTML]{DADADA}00:59:04                        \\ \cline{3-12} \cline{14-14} 
                                 & \multirow{-2}{*}{\textbf{First test case passed}}                  & \multicolumn{1}{c|}{{\ul 01:00:00}}                                          & \multicolumn{1}{c|}{00:25:05}                                                & \multicolumn{1}{c|}{{\ul 01:00:00}}                                          & \multicolumn{1}{c|}{00:29:49}                                                & \multicolumn{1}{c|}{00:49:56}                                                & \multicolumn{1}{l|}{\cellcolor[HTML]{DADADA}00:44:58}                        & \multicolumn{1}{c|}{{\color[HTML]{37638D} -}}                                & \multicolumn{1}{c|}{{\ul 01:00:00}}                                   & \multicolumn{1}{c|}{{\ul 01:00:00}}                                          & \multicolumn{1}{c|}{{\ul 01:00:00}}                                          & \multicolumn{1}{c|}{}                                                 & \cellcolor[HTML]{DADADA}00:59:04                        \\ \cline{2-12} \cline{14-14} 
\multirow{-5}{*}{\textbf{2}}     & \textbf{Submission}                                                       & \multicolumn{1}{l|}{01:00:00}                                                & \multicolumn{1}{l|}{00:31:05}                                                & \multicolumn{1}{l|}{01:00:00}                                                & \multicolumn{1}{l|}{00:45:26}                                                & \multicolumn{1}{l|}{00:50:32}                                                & \multicolumn{1}{l|}{\cellcolor[HTML]{DADADA}00:49:25}                        & \multicolumn{1}{c|}{{\color[HTML]{37638D} -}}                                & \multicolumn{1}{l|}{00:57:12}                                         & \multicolumn{1}{l|}{01:00:00}                                                & \multicolumn{1}{l|}{01:00:00}                                                & \multicolumn{1}{c|}{\multirow{-5}{*}{}}                               & \multicolumn{1}{l|}{\cellcolor[HTML]{DADADA}00:59:04}   \\ \cline{1-12} \cline{14-14} 
                                 & \textbf{First program execution}                                 & \multicolumn{1}{c|}{00:04:42}                                                & \multicolumn{1}{c|}{00:27:23}                                                & \multicolumn{1}{c|}{00:15:37}                                                & \multicolumn{2}{c|}{}                                                                                                                                       & \multicolumn{1}{l|}{\cellcolor[HTML]{DADADA}00:15:54}                        & \multicolumn{1}{c|}{00:23:31}                                                & \multicolumn{1}{c|}{00:37:04}                                         & \multicolumn{1}{c|}{00:33:19}                                                & \multicolumn{2}{c|}{}                                                                                                                                & \cellcolor[HTML]{DADADA}00:31:18                        \\ \cline{2-5} \cline{8-11} \cline{14-14} 
                                 & \cellcolor[HTML]{EFEFEF}\textbf{Interval}                        & \multicolumn{1}{l|}{\cellcolor[HTML]{EFEFEF}00:01:11}                        & \multicolumn{1}{l|}{\cellcolor[HTML]{EFEFEF}00:00:00}                        & \multicolumn{1}{l|}{\cellcolor[HTML]{EFEFEF}00:20:09}                        & \multicolumn{2}{c|}{}                                                                                                                                       & \multicolumn{1}{l|}{\cellcolor[HTML]{DADADA}00:07:07}                        & \multicolumn{1}{c|}{\cellcolor[HTML]{EFEFEF}00:01:16}                        & \multicolumn{1}{c|}{\cellcolor[HTML]{EFEFEF}00:02:07}                 & \multicolumn{1}{c|}{\cellcolor[HTML]{EFEFEF}00:01:59}                        & \multicolumn{2}{c|}{}                                                                                                                                & \cellcolor[HTML]{DADADA}00:01:47                        \\ \cline{2-5} \cline{8-11} \cline{14-14} 
                                 & \textbf{First test case passed}                                    & \multicolumn{1}{c|}{00:05:53}                                                & \multicolumn{1}{c|}{00:27:23}                                                & \multicolumn{1}{c|}{00:35:46}                                                & \multicolumn{2}{c|}{}                                                                                                                                       & \multicolumn{1}{l|}{\cellcolor[HTML]{DADADA}00:23:01}                        & \multicolumn{1}{c|}{00:24:47}                                                & \multicolumn{1}{c|}{00:39:11}                                         & \multicolumn{1}{c|}{00:35:18}                                                & \multicolumn{2}{c|}{}                                                                                                                                & \cellcolor[HTML]{DADADA}00:33:05                        \\ \cline{2-5} \cline{8-11} \cline{14-14} 
\multirow{-4}{*}{\textbf{3}}     & \textbf{Submission}                                                       & \multicolumn{1}{l|}{00:59:25}                                                & \multicolumn{1}{l|}{00:56:15}                                                & \multicolumn{1}{l|}{01:00:00}                                                & \multicolumn{2}{c|}{\multirow{-4}{*}{}}                                                                                                                     & \multicolumn{1}{l|}{\cellcolor[HTML]{DADADA}00:58:33}                        & \multicolumn{1}{l|}{00:24:50}                                                & \multicolumn{1}{l|}{01:00:00}                                         & \multicolumn{1}{l|}{01:00:00}                                                & \multicolumn{2}{c|}{\multirow{-4}{*}{}}                                                                                                              & \multicolumn{1}{l|}{\cellcolor[HTML]{DADADA}00:48:17}   \\ \hline
\end{tabular}

\label{speed-algorithm}
\end{table*}

\begin{table*}[!h]
\centerline{Table 8. Comparison of timestamps for the library-related task}
\fontsize{8.4pt}{\baselineskip}\selectfont 
\tabcolsep=0.28cm
\renewcommand\arraystretch{1.2}

\begin{tabular}{|c|c|cccccccc|}
\hline
\multicolumn{1}{|c|}{}                                &                                           & \multicolumn{8}{c|}{\textbf{Timestamps (hh:mm:ss) / Time interval (hh:mm:ss)}}                                                                                                                                                                                                                                                                                                                                                                                                          \\ \cline{3-10} 
\multicolumn{1}{|c|}{\multirow{-2}{*}{\textbf{Tasks}}} & \multirow{-2}{*}{\textbf{Key process points}}                        & \multicolumn{3}{c|}{\textbf{ChatGPT group}}                                                                                                                           & \multicolumn{1}{c|}{\cellcolor[HTML]{DADADA}\textit{\textbf{Avg}}} & \multicolumn{3}{c|}{\textbf{Stack Overflow group}}                                                                                                      & \cellcolor[HTML]{DADADA}\textit{\textbf{Avg}}         \\ \hline
                                                      & \textbf{Library installation}             & \multicolumn{1}{c|}{00:45:21}                         & \multicolumn{1}{c|}{00:53:04}                         & \multicolumn{1}{c|}{00:23:42}                         & \multicolumn{1}{c|}{\cellcolor[HTML]{DADADA}00:38:23}              & \multicolumn{1}{c|}{{\ul 01:00:00}}            & \multicolumn{1}{c|}{{\ul 01:00:00}}            & \multicolumn{1}{c|}{00:12:59}                         & \cellcolor[HTML]{DADADA}00:44:20                      \\ \cline{2-10} 
                                                      & \cellcolor[HTML]{EFEFEF}\textbf{Interval} & \multicolumn{1}{c|}{\cellcolor[HTML]{EFEFEF}00:14:39} & \multicolumn{1}{c|}{\cellcolor[HTML]{EFEFEF}00:03:44} & \multicolumn{1}{c|}{\cellcolor[HTML]{EFEFEF}00:00:08} & \multicolumn{1}{c|}{\cellcolor[HTML]{DADADA}00:06:10}              & \multicolumn{1}{c|}{\cellcolor[HTML]{EFEFEF}-} & \multicolumn{1}{c|}{\cellcolor[HTML]{EFEFEF}-} & \multicolumn{1}{c|}{\cellcolor[HTML]{EFEFEF}00:00:15} & \cellcolor[HTML]{DADADA}00:00:15                      \\ \cline{2-10} 
                                                      & \textbf{First program execution}          & \multicolumn{1}{c|}{{\ul 01:00:00}}                   & \multicolumn{1}{c|}{00:56:48}                         & \multicolumn{1}{c|}{{\color[HTML]{333333} 00:23:50}}  & \multicolumn{1}{c|}{\cellcolor[HTML]{DADADA}00:46:53}              & \multicolumn{1}{c|}{{\ul 01:00:00}}            & \multicolumn{1}{c|}{{\ul 01:00:00}}            & \multicolumn{1}{c|}{00:13:14}                         & \cellcolor[HTML]{DADADA}00:44:25                      \\ \cline{2-10} 
                                                      & \cellcolor[HTML]{EFEFEF}\textbf{Interval} & \multicolumn{1}{c|}{\cellcolor[HTML]{EFEFEF}-}        & \multicolumn{1}{c|}{\cellcolor[HTML]{EFEFEF}00:00:00} & \multicolumn{1}{c|}{\cellcolor[HTML]{EFEFEF}00:00:00} & \multicolumn{1}{c|}{\cellcolor[HTML]{DADADA}00:00:00}              & \multicolumn{1}{c|}{\cellcolor[HTML]{EFEFEF}-} & \multicolumn{1}{c|}{\cellcolor[HTML]{EFEFEF}-} & \multicolumn{1}{c|}{\cellcolor[HTML]{EFEFEF}00:00:00} & \cellcolor[HTML]{DADADA}00:00:00                      \\ \cline{2-10} 
                                                      & \textbf{First test case passed }             & \multicolumn{1}{c|}{{\ul 01:00:00}}                   & \multicolumn{1}{c|}{00:56:48}                         & \multicolumn{1}{c|}{00:23:50}                         & \multicolumn{1}{c|}{\cellcolor[HTML]{DADADA}00:46:53}              & \multicolumn{1}{c|}{{\ul 01:00:00}}            & \multicolumn{1}{c|}{{\ul 01:00:00}}            & \multicolumn{1}{c|}{00:13:14}                         & \cellcolor[HTML]{DADADA}00:44:25                      \\ \cline{2-10} 
\multirow{-6}{*}{\textbf{4}}                          & \textbf{Submission}                       & \multicolumn{1}{l|}{01:00:00}                         & \multicolumn{1}{l|}{01:00:00}                         & \multicolumn{1}{l|}{00:41:15}                         & \multicolumn{1}{l|}{\cellcolor[HTML]{DADADA}00:53:45}              & \multicolumn{1}{l|}{01:00:00}                  & \multicolumn{1}{l|}{01:00:00}                  & \multicolumn{1}{l|}{00:56:14}                         & \multicolumn{1}{l|}{\cellcolor[HTML]{DADADA}00:58:45} \\ \hline
\end{tabular}

\label{speed-library}
\end{table*}

\begin{table*}[!h]
\centerline{Table 9. Comparison of times for the two debugging tasks}
\fontsize{8.5pt}{\baselineskip}\selectfont 
\tabcolsep=0.27cm
\renewcommand\arraystretch{1.2}

\begin{tabular}{|c|c|lllllllll|}
\hline
                                                   &                               & \multicolumn{9}{c|}{\textbf{Time spent (hh:mm:ss)}}                                                                                                                                                                                                                                                                                                                                                                                                                                                                                               \\ \cline{3-11} 
\multirow{-2}{*}{\textbf{Tasks}}                   & \multirow{-2}{*}{\textbf{Bugs}}            & \multicolumn{4}{c|}{\textbf{ChatGPT group}}                                                                                                                                                       & \multicolumn{1}{c|}{\cellcolor[HTML]{DADADA}\textit{\textbf{Avg}}}           & \multicolumn{3}{c|}{\textbf{Stack Overflow group}}                                                                                                                 & \multicolumn{1}{c|}{\cellcolor[HTML]{DADADA}{\color[HTML]{333333} \textit{\textbf{Avg}}}} \\ \hline
                                                   & \textbf{Empty Stack 1}        & \multicolumn{1}{c|}{{\ul 01:00:00}} & \multicolumn{1}{c|}{00:14:25}                        & \multicolumn{1}{c|}{00:21:50}                        & \multicolumn{1}{c|}{{\color[HTML]{37638D} -}} & \multicolumn{1}{c|}{\cellcolor[HTML]{DADADA}{\color[HTML]{333333} 00:32:05}} & \multicolumn{1}{c|}{00:15:00}                        & \multicolumn{1}{c|}{00:45:28}                        & \multicolumn{1}{c|}{00:09:52}                        & \multicolumn{1}{c|}{\cellcolor[HTML]{DADADA}00:23:27}                                     \\ \cline{2-11} 
                                                   & \textbf{Division by Zero 1}          & \multicolumn{1}{c|}{{\ul 01:00:00}} & \multicolumn{1}{c|}{{\ul 01:00:00}}                  & \multicolumn{1}{c|}{{\ul 01:00:00}}                  & \multicolumn{1}{c|}{{\color[HTML]{37638D} -}} & \multicolumn{1}{c|}{\cellcolor[HTML]{DADADA}{\color[HTML]{333333} 01:00:00}} & \multicolumn{1}{c|}{{\ul 01:00:00}}                  & \multicolumn{1}{c|}{{\ul 01:00:00}}                  & \multicolumn{1}{c|}{{\ul 01:00:00}}                  & \multicolumn{1}{c|}{\cellcolor[HTML]{DADADA}01:00:00}                                     \\ \cline{2-11} 
                                                   & \textbf{Clear Stack}          & \multicolumn{1}{c|}{{\ul 01:00:00}} & \multicolumn{1}{c|}{00:33:46}                        & \multicolumn{1}{c|}{00:23:27}                        & \multicolumn{1}{c|}{{\color[HTML]{37638D} -}} & \multicolumn{1}{c|}{\cellcolor[HTML]{DADADA}{\color[HTML]{333333} 00:39:04}} & \multicolumn{1}{c|}{00:15:30}                        & \multicolumn{1}{c|}{00:45:10}                        & \multicolumn{1}{c|}{00:40:40}                        & \multicolumn{1}{c|}{\cellcolor[HTML]{DADADA}00:33:47}                                     \\ \cline{2-11} 
                                                   & \textbf{Empty Stack 2}        & \multicolumn{1}{c|}{{\ul 01:00:00}} & \multicolumn{1}{c|}{00:14:59}                        & \multicolumn{1}{c|}{{\color[HTML]{333333} 00:23:54}} & \multicolumn{1}{c|}{{\color[HTML]{37638D} -}} & \multicolumn{1}{c|}{\cellcolor[HTML]{DADADA}{\color[HTML]{333333} 00:32:58}} & \multicolumn{1}{c|}{{\color[HTML]{333333} 00:15:15}} & \multicolumn{1}{c|}{{\color[HTML]{333333} 00:45:38}} & \multicolumn{1}{c|}{{\color[HTML]{333333} 00:10:10}} & \multicolumn{1}{c|}{\cellcolor[HTML]{DADADA}00:23:41}                                     \\ \cline{2-11} 
\multirow{-5}{*}{\textbf{5}}                       & \textbf{Division by Zero 2}          & \multicolumn{1}{l|}{{\ul 01:00:00}} & \multicolumn{1}{c|}{{\ul 01:00:00}}                  & \multicolumn{1}{c|}{{\ul 01:00:00}}                  & \multicolumn{1}{c|}{{\color[HTML]{37638D} -}} & \multicolumn{1}{c|}{\cellcolor[HTML]{DADADA}{\color[HTML]{333333} 01:00:00}} & \multicolumn{1}{c|}{{\ul 01:00:00}}                  & \multicolumn{1}{c|}{{\ul 01:00:00}}                  & \multicolumn{1}{c|}{{\ul 01:00:00}}                  & \multicolumn{1}{c|}{\cellcolor[HTML]{DADADA}01:00:00}                                     \\ \hline
\multicolumn{1}{|l|}{}                             & \textbf{File Path}            & \multicolumn{1}{l|}{00:07:09}       & \multicolumn{1}{l|}{00:06:33}                        & \multicolumn{1}{l|}{00:00:31}                        & \multicolumn{1}{l|}{}                         & \multicolumn{1}{l|}{\cellcolor[HTML]{DADADA}{\color[HTML]{333333} 00:04:44}} & \multicolumn{1}{l|}{00:04:33}                        & \multicolumn{1}{l|}{00:00:39}                        & \multicolumn{1}{l|}{00:06:49}                        & \cellcolor[HTML]{DADADA}00:04:00                                                          \\ \cline{2-5} \cline{7-11} 
\multicolumn{1}{|l|}{}                             & \textbf{Set Class Index}      & \multicolumn{1}{l|}{00:29:58}       & \multicolumn{1}{l|}{00:11:13}                        & \multicolumn{1}{l|}{00:07:30}                        & \multicolumn{1}{l|}{}                         & \multicolumn{1}{l|}{\cellcolor[HTML]{DADADA}{\color[HTML]{333333} 00:16:14}} & \multicolumn{1}{l|}{00:14:20}                        & \multicolumn{1}{l|}{00:03:38}                        & \multicolumn{1}{l|}{00:45:38}                        & \cellcolor[HTML]{DADADA}00:21:12                                                          \\ \cline{2-5} \cline{7-11} 
\multicolumn{1}{|l|}{}                             & \textbf{Size of Training Set} & \multicolumn{1}{l|}{{\ul 01:00:00}} & \multicolumn{1}{l|}{{\ul 01:00:00}}                  & \multicolumn{1}{l|}{{\ul 01:00:00}}                  & \multicolumn{1}{l|}{}                         & \multicolumn{1}{l|}{\cellcolor[HTML]{DADADA}{\color[HTML]{333333} 01:00:00}} & \multicolumn{1}{l|}{{\ul 01:00:00}}                  & \multicolumn{1}{l|}{{\ul 01:00:00}}                  & \multicolumn{1}{l|}{00:36:41}                        & \cellcolor[HTML]{DADADA}00:52:14                                                          \\ \cline{2-5} \cline{7-11} 
\multicolumn{1}{|l|}{}                             & \textbf{Numeric to Normal}    & \multicolumn{1}{l|}{00:40:11}       & \multicolumn{1}{l|}{00:46:55}                        & \multicolumn{1}{l|}{00:36:35}                        & \multicolumn{1}{l|}{}                         & \multicolumn{1}{l|}{\cellcolor[HTML]{DADADA}{\color[HTML]{333333} 00:41:14}} & \multicolumn{1}{l|}{00:40:46}                        & \multicolumn{1}{l|}{{\color[HTML]{333333} 00:58:00}} & \multicolumn{1}{l|}{00:54:48}                        & \cellcolor[HTML]{DADADA}00:51:11                                                          \\ \cline{2-5} \cline{7-11} 
\multicolumn{1}{|l|}{}                             & \textbf{Remove Codes}         & \multicolumn{1}{l|}{{\ul 01:00:00}} & \multicolumn{1}{l|}{{\ul 01:00:00}}                  & \multicolumn{1}{l|}{00:21:58}                        & \multicolumn{1}{l|}{}                         & \multicolumn{1}{l|}{\cellcolor[HTML]{DADADA}{\color[HTML]{333333} 00:47:19}} & \multicolumn{1}{l|}{{\ul 01:00:00}}                  & \multicolumn{1}{l|}{00:27:47}                        & \multicolumn{1}{l|}{01:00:00}                        & \cellcolor[HTML]{DADADA}00:49:16                                                          \\ \cline{2-5} \cline{7-11} 
\multicolumn{1}{|c|}{\multirow{-6}{*}{\textbf{6}}} & \textbf{Accuracy}             & \multicolumn{1}{l|}{{\ul 01:00:00}} & \multicolumn{1}{l|}{{\color[HTML]{333333} 00:54:57}} & \multicolumn{1}{l|}{{\color[HTML]{333333} 00:39:02}} & \multicolumn{1}{l|}{\multirow{-6}{*}{}}       & \multicolumn{1}{l|}{\cellcolor[HTML]{DADADA}{\color[HTML]{333333} 00:51:20}} & \multicolumn{1}{l|}{00:41:58}                        & \multicolumn{1}{l|}{00:58:00}                        & \multicolumn{1}{l|}{01:00:00}                        & \cellcolor[HTML]{DADADA}00:53:19                                                          \\ \hline
\end{tabular}

\label{speed-debug}
\end{table*}

\subsection{RQ3: Reasons for Differential Effects of ChatGPT and Stack Overflow}

\begin{table}[]
\centerline{Table 10. Comparison of ratings between the two platforms}
\fontsize{7.8pt}{\baselineskip}\selectfont 
\tabcolsep=0.06cm
\renewcommand\arraystretch{1.1}
\centering
\begin{tabular}{|c|c|c|c|c|c|
>{\columncolor[HTML]{DADADA}}c |}
\Xhline{1.5px}
\textbf{Platforms}                     & \textbf{1s count} & \textbf{2s count} & \textbf{3s count} & \textbf{4s count} & \textbf{5s count} & \textbf{Avg score} \\ \Xhline{1.5px}
\multicolumn{1}{|c|}{\textbf{ChatGPT}} & 0                       & 2                            & 5                              & 7                         & 9                        & 4.0               \\ \hline
\textbf{Stack Overflow}                & 3                       & 8                            & 7                              & 3                         & 0                        & 2.5               \\ \Xhline{1.5px}
\end{tabular}
\end{table}

\begin{table*}[]
\centerline{Table 11. Strengths and weaknesses of ChatGPT and Stack Overflow analyzed from the post-experiment questionnaires}
\fontsize{7.8pt}{\baselineskip}\selectfont 
\tabcolsep=0.1cm
\renewcommand\arraystretch{1.1}
\begin{tabularx}{1.0\textwidth}{|c|c|c|X|}
\Xhline{1.5px}
\multicolumn{1}{|c|}{\textbf{Platforms}}  &                                  \textbf{Aspects}              &    \textbf{S/W}                                &     \textbf{Key points with frequencies and examples of original participant responses}                                                                                                                                                                                                                                                                                                         \\ \Xhline{1.5px}
\multirow{10}{*}{\textbf{ChatGPT}}        & \multirow{4}{*}{\textbf{Quality of answers}} & \multirow{2}{*}{\textbf{Strengths}}  & \textbf{details to write code (6):} ``Assistance in syntax, methods, or other technical details that I cannot remember."                                                                                                                                                                      \\ \cline{4-4} 
                                          &                                              &                                      & \textbf{provide algorithm templates (6):} ``ChatGPT can provide algorithms for classic problems."                                                                                                                                                                                             \\ \cline{3-4} 
                                          &                                              & \multirow{2}{*}{\textbf{Weaknesses}} & \textbf{wrong and expired answers (25):} ``Links given are expired."; ``The code is wrong."                                                                                                                                                                                                   \\ \cline{4-4} 
                                          &                                              &                                      & \textbf{cannot handle uncommon problems (1):} ``At present, it still lacks the ability to create new algorithms and can only use known algorithms, making it difficult to handle some rare problems."                                                                                         \\ \cline{2-4} 
                                          & \multirow{6}{*}{\textbf{User experiences}}   & \multirow{4}{*}{\textbf{Strengths}}  & \textbf{provide ideas (15):} ``It can provide ideas to solve the problem exactly."                                                                                                                                                                                                            \\ \cline{4-4} 
                                          &                                              &                                      & \textbf{generate and explain code (16): } ``It can generate code and give a clear structure."                                                                                                                                                                                                 \\ \cline{4-4} 
                                          &                                              &                                      & \textbf{help debug (3):} ``Returning samples to GPT would enable it to perform automatic error correction."                                                                                                                                                                                   \\ \cline{4-4} 
                                          &                                              &                                      & \textbf{help describe questions (1):} ``ChatGPT possesses strong text comprehension ability, and sometimes users may not be fully aware of the problems they need to search. However, ChatGPT can filter out related questions that users may be interested in while answering the question." \\ \cline{3-4} 
                                          &                                              & \multirow{2}{*}{\textbf{Weaknesses}} & \textbf{need to understand code for code refinement (7):} ``Since the code was not written by me, it requires understanding before debugging."                                                                                                                                                                       \\ \cline{4-4} 
                                          &                                              &                                      & \textbf{may mislead (2):} ``The code does not match the results of the test cases it suggests, which can cause confusion and be misleading."                                                                                                                                                  \\ \hline
\multirow{12}{*}{\textbf{Stack Overflow}} & \textbf{Number of questions}                 & \textbf{Weaknesses}                  & \textbf{searchable questions are not enough (5):} ``I can only find answers to solved problems."                                                                                                                                                                                              \\ \cline{2-4} 
                                          & \multirow{2}{*}{\textbf{Number of answers}}  & \textbf{Strengths}                   & \textbf{extra knowledge (3):} ``Some answers may expand on related knowledge."                                                                                                                                                                                                                 \\ \cline{3-4} 
                                          &                                              & \textbf{Weaknesses}                  & \textbf{answers are not enough (11):} ``The assistance for basic syntax is weak.""Unable to find the desired template."                                                                                                                                                                       \\ \cline{2-4} 
                                          & \multirow{5}{*}{\textbf{Quality of answers}} & \multirow{2}{*}{\textbf{Strengths}}  & \textbf{solve explicit exceptions (3):} ``Solutions to exceptions can be easily obtained."                                                                                                                                                                                                    \\ \cline{4-4} 
                                          &                                              &                                      & \textbf{useful links (7):} ``Compared to the answers themselves, the GitHub links or Apache Maven repository instructions provided are more useful."                                                                                                                                          \\ \cline{3-4} 
                                          &                                              & \multirow{3}{*}{\textbf{Weaknesses}} & \textbf{few and scattered codes (1):} ``The searched codes are fragmented and cannot be easily combined due to incompatibility."                                                                                                                                                              \\ \cline{4-4} 
                                          &                                              &                                      & \textbf{bad answers (3):} ``The code may contain errors and it can be difficult to determine."                                                                                                                                                                                                \\ \cline{4-4} 
                                          &                                              &                                      & \textbf{no detailed explanation (6):} ``Some solutions to the problem may not have detailed explanations and contextual information."                                                                                                                                                         \\ \cline{2-4} 
                                          & \multirow{4}{*}{\textbf{User experiences}}   & \textbf{Strengths}                   & \textbf{communication (1):} ``It provides a platform to communicate."                                                                                                                                                                                                                         \\ \cline{3-4} 
                                          &                                              & \multirow{3}{*}{\textbf{Weaknesses}} & \textbf{cannot provide ideas (6):} ``It is of no help in some algorithmic problems."                                                                                                                                                                                                          \\ \cline{4-4} 
                                          &                                              &                                      & \textbf{inapposite order of answers (2):} ``The answer I need may be ranked relatively low in the search results."                                                                                                                                                                            \\ \cline{4-4} 
                                          &                                              &                                      & \textbf{need search skills (5):} ``The search terms used may not be precise enough, resulting in answers that do not apply."                                                                                                                                                                  \\ \Xhline{1.5px}
\end{tabularx}
\end{table*}

To investigate the differing performance of ChatGPT and Stack Overflow in helping programmers generate high-quality code and complete tasks quickly, we conducted a post-experiment questionnaire survey with 44 participants. Of these, 23 were in the ChatGPT group, and 21 were in the Stack Overflow group. In the questionnaire, participants were asked to rate chatGPT (or Stack Overflow) and provide their perceptions of its strengths and weaknesses. It should be noted that a higher score indicates that participants perceive the tool to be more helpful.

We calculated the frequency of ratings from 1 to 5 for the two platforms, the results are presented in Table 10.  In the table, ``1s count" refers to the total number of participants who rated with a score of 1, and columns 3 to 6 follow the same interpretation. The ``Avg score" column calculates the average scores for both platforms, with ChatGPT scoring 4.0 and Stack Overflow scoring 2.5. ChatGPT's average score is significantly higher than that of Stack Overflow.

We processed the responses to the two open-ended questions regarding the strengths and weaknesses of ChatGPT (or Stack Overflow) in the following manner: (1) we extracted key points from each participant's response; (2) for each platform's strengths and weaknesses, we grouped similar key points together and calculated their frequencies. Note that the total count of key points may not match the number of participants due to multiple viewpoints expressed in the questionnaire.
(3) starting from the features of ChatGPT and Stack Overflow, we divided users' feedback on them into 2 and 4 main aspects respectively, and classified the key points obtained into these main aspects. Since the usage of ChatGPT is to provide a demand and then get an answer, we divided users' feedback on it into two main aspects: answer quality and user experience. Since the usage of Stack Overflow is to search for a question and then find a suitable answer to get solution, we divided users' feedback on it into four main aspects: number of questions, number of answers, answer quality, and user experience. Notice that the analysis process of this questionnaire was carried out through multiple discussions among the three authors among us to reach a consensus.

Table 11 presents a comparison of the strengths and weaknesses of the two platforms. We selected the top two key points with the highest frequencies, which we consider to represent points of consensus among participants. Our findings reveal that participants perceived the key strengths of ChatGPT to be ``generate and explain code" (16) and ``provide ideas" (15) while its weaknesses include ``wrong and expired answers" (25) and ``need to understand code for code refinement" (7). As for Stack Overflow, its key strengths encompass ``useful links" (7), ``extra knowledge" (3), and ``solve explicit exceptions" (3), while its weaknesses involve ``answers are not enough" (11), ``no detailed explanation" (6), and ``cannot provide ideas" (6). In comparing code quality, ChatGPT outperforms Stack Overflow in algorithmic and library-related tasks. This is because participants can easily generate code and obtain insights from ChatGPT. On the other hand, Stack Overflow lacks clear questions and answers, making it difficult for participants to gain problem-solving insights. This also explains why ChatGPT shows a significantly faster completion time in algorithmic tasks compared to Stack Overflow. For debugging tasks, Stack Overflow is better due to its ability to solve explicit exceptions and provide useful links for resolution methods, which ChatGPT lacks. The reason why ChatGPT users requires more time for code refinement may be there are errors and outdated information included in its answers.


We also made an interesting observation where 3 participants mentioned that using Stack Overflow allowed them to gain more knowledge during the problem-solving process (``extra knowledge'' key point in Table 11), 7 participants complained about the need to understand the generated code from ChatGPT (``need to understand code for code refinement'' key point in Table 11). This finding may suggest that excessive reliance on ChatGPT could potentially limit our problem-solving thought, thereby diminishing certain aspects of creativity. This could also be a potential reason why Stack Overflow outperforms ChatGPT in debugging tasks.

\begin{tcolorbox}[boxrule=1pt,boxsep=1pt,left=2pt,right=2pt,top=2pt,bottom=2pt]
\textbf{Answer to RQ3:} 
ChatGPT is obviously better than Stack Overflow in algorithm and library tasks due to the fact that ChatGPT can quickly generate code and provide ideas, while Stack Overflow lacks task-related questions and answers.
Stack Overflow has an advantage in debugging tasks due to its expertise in solving explicit exceptions and providing helpful links.
\end{tcolorbox}

\begin{tcolorbox}[boxrule=1pt,boxsep=1pt,left=2pt,right=2pt,top=2pt,bottom=2pt]
\textbf{Observation 3:} 
Excessive reliance on ChatGPT may potentially limit our problem-solving thought, thereby diminishing certain aspects of creativity.
\end{tcolorbox}

\section{Discussions}



\subsection{Threats to Validity}


\textbf{Assessment of participants' programming abilities:} In the comparative experiment, we evaluated the participants' programming abilities through survey results and discussions with the authors, which may pose a threat to the validity of the experiment. In order to minimize this threat, we investigated the participants' abilities in multiple aspects in the questionnaire and the three authors reached a consensus through discussion to eliminate subjective bias as much as possible.


\textbf{Random assignment of participants to either the ChatGPT or Stack Overflow group:} In the comparative experiment, the random allocation of participants into the ChatGPT and Stack Overflow groups may introduce a threat to the validity of the results due to potential differences in participants' familiarity with the two platforms. To address this potential threat, we provided each participant with guidance manuals for either ChatGPT or Stack Overflow.

\subsection{Limitations}

Our work has the following limitations: (1) The relatively small sample size of 44 participants in our user study; (2) The participants were all computer science students, which may lack representativeness for the broader programmer population; (3) The three types of programming tasks we designed may not cover all possible usage scenarios. We intend to address these limitations in our future work.

\section{Related Works}

In this section, we discuss recent related research in three aspects.

\subsection{User Studies of Stack Overflow}

As of January 2023, Stack Overflow is boasting over 23 million registered users and a vast repository of questions and answers contributed by a multitude of programmers~\cite{SO2023}. In order to assess the extent of assistance provided by Stack Overflow to programmers, numerous researchers have conducted user studies. 

Dondio et al.~\cite{SOAsPlugin} conducted a user study aimed at evaluating the impact of using Stack Overflow as a supplementary plugin to enhance the academic performance of students (N=38), it was found that the experimental group, which utilized Stack Overflow as a tutorial tool, demonstrated improved performance in the same topic compared to the control group, which used traditional learning formats. 
Lo et al.~\cite{SOLimit} discovered limitations in Stack Overflow. 
Some more advanced MATLAB topics such as Simulink, image processing, signal processing, and computer vision are less likely to be addressed on the platform. Wang et al.~\cite{badge} analyzed over 3 million revision records provided by users confirming the relationship between the badge system and users' revision of answers, and this finding has been acknowledged by the Stack Overflow community.
Fangl et al.~\cite{polite} focused on a specific politeness strategy-expressing gratitude-in Q\&A sites, finding that gratitude expressions in comments can motivate answerers to generate higher-quality content. Wijekoon et al.~\cite{SODataMining} analyzed the global user distribution and contribution
 of Stack Overflow.

\subsection{User Studies of Language Models for Code Assistance}

In recent years, the rapid development of natural language processing techniques and the widespread application of language models have provided new assistance tools such as ChatGPT and GitHub Copilot~\cite{GithubCopilot}. To assess the effectiveness and feasibility of these methods, researchers have conducted several user studies.



Perry et al.~\cite{perry2022users} conducted a user study (N=54) on AI Code assistants for security tasks, finding that users with access to the assistant were more likely to introduce security vulnerabilities but also rated their insecure answers as secure.
Sandoval et al.~\cite{sandoval2023lost} conducted a user study (N=58) on LLM code suggestions, showing that AI-assisted users produced critical security bugs at a rate no more than 10\% higher than the control group, indicating a likely beneficial impact on functional correctness and no significant increase in severe security bugs.
Vaithilingam et al.~\cite{CopilotPython} measured productivity from a group of developers (N=24) completing code tasks in Python.  They found Copilot generates code a lot quicker than typing or finding it from other sources but it is often buggy.
Imai~\cite{pair-program} tasked a group of developers (N=21) to implement code for a ‘minesweeper’ game. 
The study concluded that Copilot tended to result in more lines of code than the human-based pair-programming in the same amount of time. However, the quality of code produced by Copilot was lower. Pair-programming with Copilot does not match the profile of human pair-programming. The study did not examine whether or not Copilot improves over a developer without a pair programmer.
A study by Ziegler et al.~\cite{Productivity} from GitHub examines user perspectives on productivity during usage of GitHub Copilot.
Users felt Copilot had a more beneficial effect on their productivity than a negative one.


\subsection{Comparative Experiments of Programming Assistance Platforms}

In order to compare two platforms or tools, researchers have conducted comparative experiments. These experiments typically entail gathering data on specific problems that each platform is capable of addressing, followed by an analysis of the effectiveness and the advantages and disadvantages of the two platforms using mathematical methods.

Stack Overflow is used commonly by developers to seek answers for their software development as well as privacy-related concerns. Delile et al.~\cite{delile2023evaluating} conducted a comparative analysis between the accepted responses  given by Stack Overflow users and the responses produced by ChatGPT for those extracted questions to identify if ChatGPT could serve as a viable alternative to generate code or produce  responses to developers’ questions. The findings illustrate that ChatGPT generates similarly correct responses for about 56\%  of questions while for the rest of the responses, the answers from Stack Overflow are slightly more accurate than ChatGPT. Our study differs from this article in that we focus on the different effects of platforms on enhancing programmer productivity.


\section{Conclusion}




We conducted a user study to investigate the differing performance of ChatGPT and Stack Overflow in helping programmers generate high-quality code and complete tasks quickly. We found that ChatGPT is obviously better than Stack Overflow in algorithm and library tasks, while
Stack Overflow has an advantage in debugging tasks. Additionally, we conducted a post-experiment questionnaire survey to identify the reasons behind the two platforms’ divergent performances. In future research, we will try to address the limitations introduced in Section 5.2.


\begin{thebibliography}{10}
\bibitem{maurergeo}
B.~Maurer, ``Geo p. tech, ai chatbot geotechnical engineer: How ai language
  models like “chatgpt” could change the profession,''
\end{thebibliography}
\end{document}